\documentclass[%
reprint,
 amsmath,amssymb,
 aps,
]{revtex4-1}
\usepackage{graphicx,subfigure,color}
\usepackage{dcolumn}
\usepackage{bm}
\usepackage{hyperref}
\usepackage[mathlines]{lineno}
 
\newcommand{\be}{\begin{equation}}
\newcommand{\ee}{\end{equation}}

\begin{document}

\title{Strong Correlation Effects on Surfaces of Topological Insulators via Holography}  
\author{Yunseok Seo, Geunho Song and Sang-Jin Sin}
%
 \affiliation{Department of Physics, Hanyang University, Seoul 04763, Korea.}  \date{\today}%
\begin{abstract}
We investigate   effects of strong correlation on the surface state of topological insulator (TI). 
We   argue that electrons in the regime of crossover from weak anti-localization to weak localization, are strongly correlated and calculate  magneto-transport coefficients of TI  using gauge gravity principle. Then, we examine,  magneto-conductivity (MC) formula and find excellent agreement with the data of chrome doped Bi$_{2}$Te$_{3}$ in the  crossover regime. 
We also find that cusp-like peak in MC at low doping is absent, 
which is natural since quasi-particles disappear due to the strong correlation. 
\begin{description} 
\item[PACS numbers]11.25.Tq, 71.10.-d, 72.15.Rn
\end{description}
\end{abstract}
\pacs{here} 
\keywords{Surface state,  topological  insulator, holography }
\maketitle

{\bf Introduction:}
Understanding strongly correlated  electron systems  has been a theoretical challenge  for several decades. 
Typically, such systems lose quasi-particles  and  show mysteriously  rapid thermalization \cite{Sachdev:2011mz,zaanen2015holographic,Oh:2013qxn,Sin:2013yha}, which provide the hydrodynamic description \cite{HKMS,Lucas:2015sya} of them near quantum critical point (QCP).  
Recently, the principle of gauge-gravity duality \cite{Maldacena:1997re,Witten:1998qj,Gubser:1998bc}  attracted much   interest  as a possibility of the paradigm for strongly interacting systems, where the system near QCP is mapped to  a black hole.
More recently,  large violation of Widermann-Frantz law was observed in graphene near charge neutral point, indicating that it is 
a strongly   interacting system \cite{pkim}  in a window of temperature, 
 and the gauge gravity principle applied to it exhibited remarkable agreement  
with the  experimental data \cite{Seo:2016vks}. 

The fundamental reason for the appearance of the strong interaction in graphene is the 
{\it  smallness of the fermi sea}: in the presence of the Dirac cone,  when fermi surface is near the tip of the cone, electron hole pair creation from such a small fermi sea is insufficient to screen the Coulomb interaction. 
Because  this is so simple and universal,  one  can expects that for any Dirac material, there should be a  regime of parameters where electrons are strongly correlated. Dirac cone  also provides the reason why it is a quantum critical system with Lorentz invariance. 
The most well known Dirac material other than the graphene is the surface of a topological insulator (TI) \cite{hasan2010colloquium,qi2011topological}. The latter has an unpaired Dirac cone and strong spin-orbit coupling, and   as a consequence, it has a  variety of interesting  physics\cite{Yu61,PhysRevLett.102.216403,PhysRevB.78.195424} including weak anti-localization (WAL) \cite{bergmann1982weak}. 

Magnetic doping in TI can open  a gap in the  surface state  by  breaking the time reversal  symmetry \cite{liu2012crossover,zhang2012interplay,bao2013quantum},
and it is responsible  for the transition from WAL to weak localization(WL). 
For extreme low doping,  the  sharp horn of the magneto-conductivity curve  near zero magnetic field can be   described by  Hikami-Larkin-Nagaoka (HLN) function \cite{hikami1980spin}.
However, for intermediate doping  where the tendency of WAL and weak localization (WL) compete,   a satisfactory theory is still wanted \cite{lu2011competition,liu2012crossover,bao2013quantum} although there is a phenomenological description  \footnote{In ref. \cite{liu2012crossover,lang2012competing},  
the authors   assigned  weights  for two HLN functions of opposite sign by hand to fit the  data.  
In graphene case,   such transitions are 
better understood. \cite{mccann2006weak,tikhonenko2009transition} in terms of inter-valley scattering versus spin-orbit interaction. } 
Even in the case the fermi surface is large at low doping so that the system is a fermi liquid,  
increasing the surface gap pushes up the dispersion curve, which makes the fermi sea small.
Then,   the   logic for strong coupling in graphene works  for transition region in surface of TI. 
Therefore electron system near the transition region should be strongly correlated. 

In this paper, we investigate magneto-conductivity (MC)  for  the surface  of  a topological insulator with  correlated electrons using gauge gravity principle.
We   will give  analytic formulae  of all the magneto-transports  on the surface 
of TI with strong correlation  as a function of  magnetic field, temperature and impurity  density   and compare the result with   Bi$_{2}$Te$_{3}$ data of \cite{bao2013quantum}.   
Most interestingly, in the doping regime with crossover from WAL to WL,  
our theory agrees with experimental data nicely in a window of temperature justifying our suggestion that   electrons in the experimented  material  are strongly correlated in  this regime.  
Our results also show that  the cusp-like peak  in  MC curve at fixed temperature, which is the hall-mark of WAL in  the weakly interacting system, is absent, which can be   argued   to be a  consequence of strong correlation. 

{\bf Idea of the model:} 
Our system is the surface  of topological insulator which is a 2+1 dimensional system with odd number of Dirac cones. Our question is what happens if such system has strong correlation as well  and  the recipe for strong electron-electron interaction is to use gauge gravity principle or holography. For TI, special care is necessary to encode strong spin-orbit coupling (SOC).  
Our holographic model is defined on a manifold $\mathcal{M}$  which is asymptotically AdS$_4$. 
With these setup, our model is defined by the action,  
\begin{align}\label{action}
	2\kappa^2S&=\int_{\mathcal{M}} d^4x \sqrt{-g}\left [R+\frac{6}{L^2}-\frac{1}{4}F^2-\sum_{I,a=1,2}\frac{1}{2}(\partial\chi_I^{(a)})^2\right] \nonumber \\
	&\quad \quad\quad -\frac{q_{\chi}}{16}\int_{\mathcal{M}}\sum_{I=1,2}(\partial\chi_{I}^{(2)})^2 F\wedge F 
\end{align}
where $q_{\chi}$ is the coupling   and $\kappa^2=8\pi G$ and $L$ is the AdS radius. From now on,  we set $2\kappa^2=L=1$.  
The action contains  two pairs of  bosons, one for the magnetic impurities and the other for the  non-magnetic ones.  To  encode the effect of SOC in the presence of the magnetic doping,  
we introduced  the last term which is a  coupling between the  impurity density and the instanton density. 
Such an interaction term  was first  introduced  in \cite{kkss} by us to discuss the SOC. 
The strong SOC  provides the band inversion that induces massless chiral fermions at the boundary, which in turn induces the chiral anomaly as a nontrivial divergence of the chiral current. 
In fact, our interaction term is  unique  in that it is the leading order term that  can  take care of  
  anomaly and its coupling to  impurity in a manner with time reversal symmetry   broken. 

The solution of equation of motion is given by  
 \begin{align}\label{bgsol0}
 	&A=a(r)dt+\frac{1}{2}H(xdy-ydx),\nonumber\\
 	&\chi_I^{(1)}=\alpha \left(\begin{array}{c}    x \\   y \end{array}  \right),\qquad \chi_I^{(2)}=\lambda \left(\begin{array}{c}   x \\      y \end{array}  \right),\nonumber\\
 	&ds^2=-U(r)dt^2+\frac{dr^2}{U(r)}+r^2(dx^2+dy^2),
 \end{align}  
 \begin{align}\label{bgsol}
 {\rm with }\quad	&U(r)=r^2-\frac{\alpha^2+\lambda^2}{2}-\frac{m_0}{r}+\frac{q^2+H^2}{4r^2}\cr
	&~~~~~~~~+\frac{\lambda^4H^2q_{\chi}^2}{20r^6}-\frac{\lambda^2Hqq_{\chi}}{6r^4},\cr
 	&a(r)=\mu-\frac{q}{r}+\frac{\lambda^2Hq_{\chi}}{3r^3},
 \end{align}
 where $\mu$ is the chemical potential, $q$ is charge carrier density. $q$ and $m_0$ is determined by    the regularity condition  at the black hole horizon, $A_t(r_0)=U(r_0)=0$. 
 \begin{align}\label{m0q}
 	q&=\mu r_0 +\frac{1}{3}\theta H \qquad \text{with} \quad \theta=\frac{\lambda^2q_{\chi}}{r_0^2}\\
 	m_0&=r_0^3\left(1+\frac{r_0^2\mu^2+H^2}{4r_0^4}-\frac{\alpha^2+\lambda^2}{2r_0^2}\right)+\frac{\theta^2H^2}{45r_0}.
 \end{align}
Usually the chemical potential is proportional to the charge density.  However,  in our model, there is extra term   $\sim \theta H$, which represents  the Witten effect, the magnetic field generation by  electric charge and vice versa. 
It comes from the last term of the action whose microscopic  origin is the spin-orbit interaction\cite{nagaosa2010anomalous,essin2009magnetoelectric}. 
\vskip.3cm 


 The temperature of the physical system is identified by the Hawking temperature of the black hole given by 
\begin{align}\label{temperature}
T = \frac{12r_0^4 - \left[ H^2 +2 r_0^2 (\alpha^2+\lambda^2) +(q -H \theta)^2   \right]} {16\pi  r_0^3}
\end{align}
and the entropy and energy densities are  given by  
$s= 4\pi r_0^2$, $ \epsilon = 2 m_0$ respectively.
Since the boundary on-shell action is related with 
 pressure by  $S_{onshell}=-{\cal P}$, we   get   
$\varepsilon +{\cal P} = s \, T + \mu\, q.$
Then, the magnetization can be obtained from 
$M = -\frac{\partial \epsilon}{\partial H}.$
\vskip.3cm

\noindent {\bf   DC transport coefficients} can be calculated by 
 turning  on small fluctuations around background   (\ref{bgsol}) \cite{Donos:2014cya};
\begin{align}
\delta G_{ti} &= -t U(r) \zeta_i +\delta g_{ti} (r), \quad 
\delta G_{ri} =  r^2 \delta g_{ri} \cr
\delta A_i  &= t (-E_i + \zeta_i a(r) ) + \delta a_i (r),
\end{align}
where $i =x$, $y$. 
Notice that equations of motion for fluctuation are time-independent, although there is explicit time dependence in above ansatz.  
Here $E_{i}$ corresponds to the external electric field and $\zeta_i=-\partial_{i} T/T$. We   define bulk currents by 
\begin{align}\label{current0}
{\cal J}^{i} =\sqrt{-g} F^{ir},\;
{\cal Q}^{i}=U(r)^2 \partial_r \left(\frac{\delta g_{ti}(r)}{U(r)}\right) - a_t (r) J^i.
\end{align}
which    become the  electric and the heat current   $J^{i}$, $Q^{i} = \langle  T^{ti}\rangle - \mu J^{i}$ respectively at the boundary($r\rightarrow \infty$). 
Using  the equations of motion of the fluctuation fields  together with the horizon regularity condition, we can get  electric and heat current at the boundary in terms of the external sources;
\begin{align}\label{current3}
J^{i} =& \frac{({\cal F}+{\cal G}^2)({\cal F}-H^2)}{{\cal F}^2 + H^2 {\cal G}^2} E_i \cr
& +\left[\theta +\frac{H {\cal G}(2 {\cal F} +{\cal G}^2 -H^2)}{{\cal F}^2 + H^2 {\cal G}^2}  \right] \epsilon_{ij} E_j \cr
&+ \frac{s T {\cal G}({\cal F}-H^2)}{{\cal F}^2 + H^2 {\cal G}^2} \zeta_i +\frac{s T H({\cal F}+{\cal G}^2)}{{\cal F}^2 + H^2 {\cal G}^2} \epsilon_{ij} \zeta_j \cr
Q^{i} =& \frac{s T {\cal G}({\cal F}-H^2)}{{\cal F}^2 + H^2 {\cal G}^2} E_{i} +\frac{s T H({\cal F} +{\cal G}^2)}{{\cal F}^2 + H^2 {\cal G}^2} \epsilon_{ij} E_{j} \cr
&+\frac{s^2 T^2 {\cal F}}{{\cal F}^2 + H^2 {\cal G}^2} \zeta_{i} +\frac{s^2 T^2 H {\cal G}}{{\cal F}^2 + H^2 {\cal G}^2}\epsilon_{ij} \zeta_{j}  ,
\end{align}
where  $\zeta_{i}=-(\nabla_i T)/T $ as before and 
\begin{align}
{\cal F} &= r_0^2 (\alpha^2+\lambda^2) +(1+\theta^2) H^2 -q\, \theta H \cr
{\cal G} &=q -\theta H. 
\end{align}

Now, the transport coefficients can be read off from
\begin{align}\label{transport}
\left(\begin{array}{c} J^{i} \\ Q^{i} \end{array}  \right)=\left(\begin{array}{cc} \sigma_{ij} & \alpha_{ij} T \\ \bar{\alpha}_{ij} T& \bar{\kappa}_{ij} T \end{array}\right) \left( \begin{array}{c} E_j \\ \zeta_{j} \end{array}   \right).
\end{align}
In $q_{\chi} \to 0$ limit,   Eq. (\ref{current3}) are reduced to  those of dyonic black hole \cite{Andrade:2013gsa,Amoretti:2015gna,Blake:2015ina,Kim:2015wba}. 
There are two important symmetries of the DC conductivities: one is the
anti-symmetry  of the  off-diagonal components, i.e,  $X_{ij}=-X_{ji}$ for all $X=\sigma, \alpha,{\bar \kappa}$;    
and the other is $\alpha_{ij}$=$\bar{\alpha}_{ij}$,  which is  Onsager's relation. 
If we further take $H \to 0$ limit,  \begin{align}
\sigma_{xx}  \to  1+ \frac{q^2}{r_0^2(\alpha^2 +\lambda^2)}. \label{sold}
\end{align}
Notice that if we define 
$\beta^2 = \alpha^2 +\lambda^2$,~$\gamma = \frac{\lambda^2}{\alpha^2 +\lambda^2}$, then 
$\beta^{2}$ plays the role of the total impurity density used in \cite{kkss},  and  $\lambda^2$ and $ \alpha^2$ can be interpreted as  the magnetic and non-magnetic impurity density respectively.  
Therefore  $\gamma$ corresponds to the magnetic doping parameter, which is usually denoted by x in the literature.  
 \vskip .1cm 
  
  \noindent
{\bf Magneto-conductance:} 

To compare our results with  the data for the non-ferro magnetic material, we take $\mu = 0$
to set  the ferromagnetic magnetization zero. 
The longitudinal conductivity in this limit  is 
\begin{align}\label{DC01}
\sigma_{xx} &= \frac{({\cal F}+{\cal G}^2)({\cal F}-H^2)}{{\cal F}^2 + H^2 {\cal G}^2}.
\end{align}
The MC  is defined by 
$\Delta \sigma \equiv \sigma_{xx} (H) -\sigma_{xx} (0)$.

Consider the evolution of the system  with   the doping. As the
surface gap  increases, the size of the fermi surface decreases.  
See figure 1(a).  At x=0.08 gap is large enough to see transition from WAL to WL for some temperature, but fermi surface is still large so that particle character remains. At x=0.1, gap is large enough 
 and fermi surface is small enough to show  strong coupling behavior, so that our theory is well applicable.   
Figure 1(b)  shows the evolution of MC curve as we raise 
the doping rate assuming  that entire regime can be described holographically.  
However the real system is strongly correlated only when fermi surface is small enough.
Therefore we expect that our theory is valid only  in a window of doping rate as well as that of temperature. 
This is indeed what happens.  In figure 1(b),  the green color indicates the validity island in parameter space of $(\gamma, H)$, where our theory agrees with  experimental result of ref.[20]. 
  
\begin{figure}[ht!]
\centering
    \subfigure[ ]
   {\includegraphics[width=2.8cm]{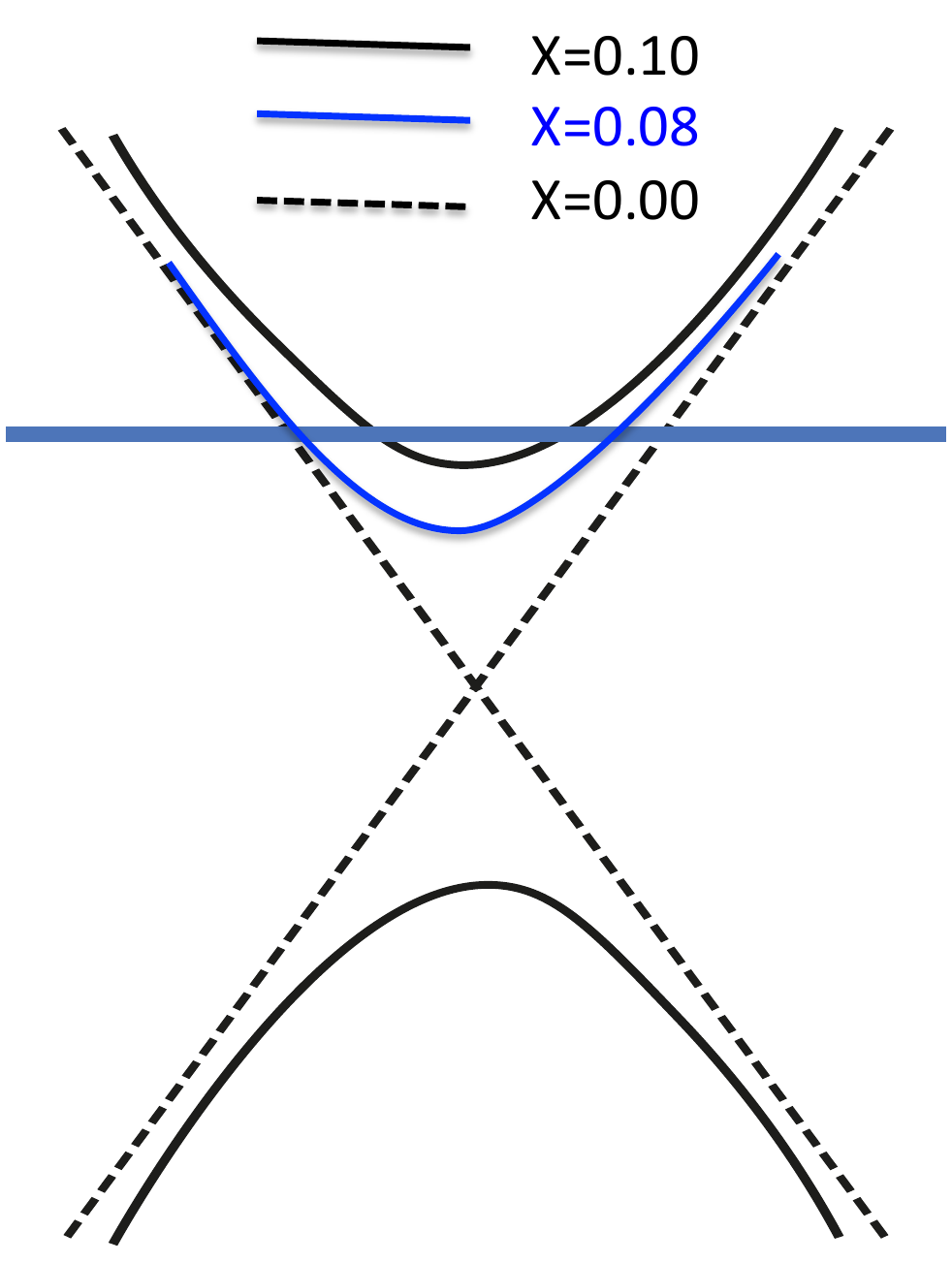}  }
\hskip.5cm   
       \subfigure[ ]
   {\includegraphics[width=4.2cm]{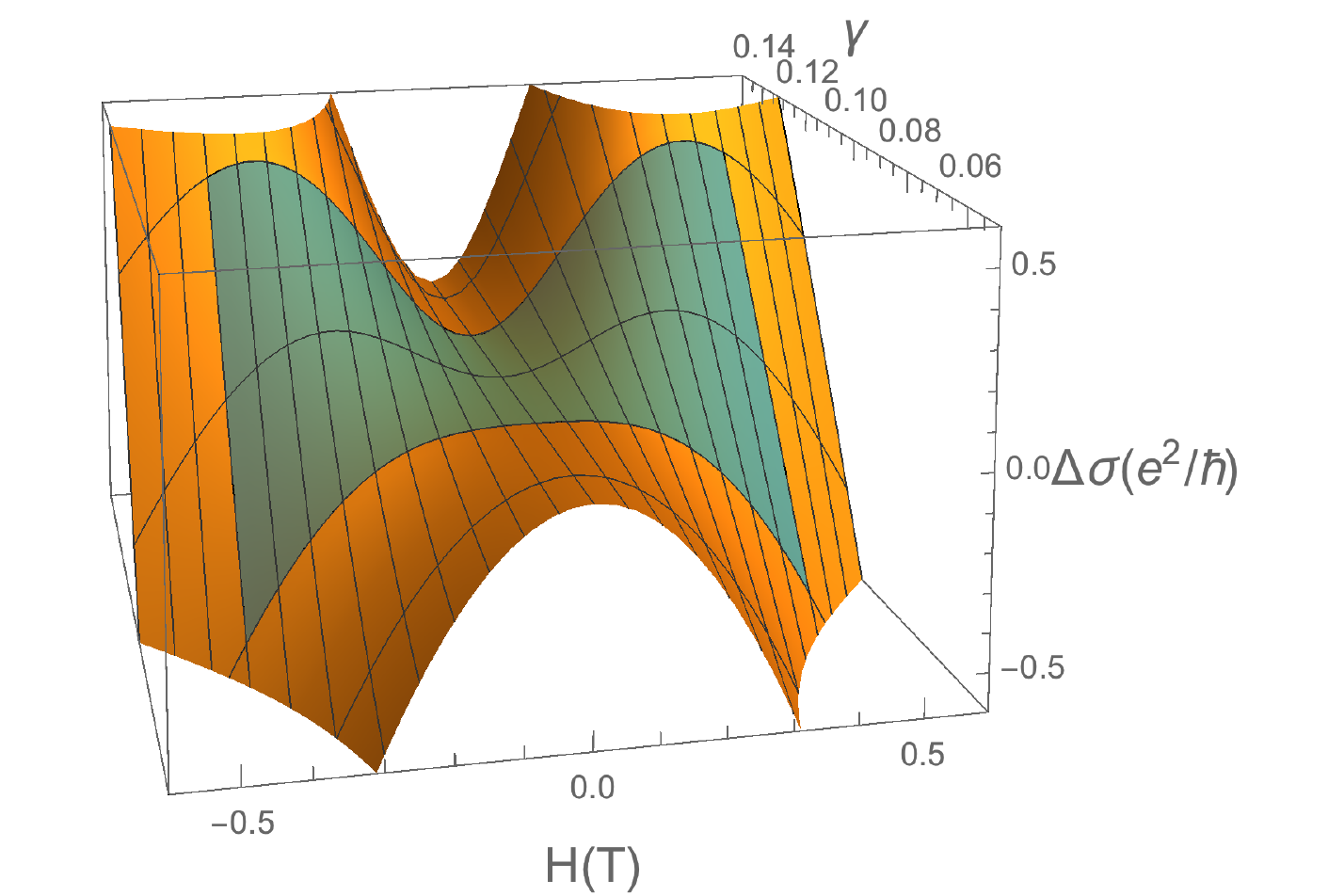}  }
    \caption{Evolution of  (a) density of state and (b) MC as we vary the doping.  
    Again, our theory fit data only in an island of parameter space $(H,\gamma)$, where $\gamma=x$.    } \label{fig:evolution}
\end{figure}
 
As we discussed earlier, the problematic part of the data fitting in weakly interacting picture is the 
medium doping regime $x\sim 0.1$ 
where the transition between the WAL to WL is smooth.  Does our theory fit data in such region? 
The answer is given in figure 2, where we took the data for   $x=0.1$. Here again, 
our theory is valid only in an island of parameter space $(H,T)$. 
 There are only 4 adjustable parameters:  $\gamma, \beta,q_ {\chi}, v_{F}$. 
 Others ($T,H,\mu$) are  plot variables.  
From the data fitting point of view, the 1.9 K data is difficult to fit   because it has too steep curvature  near zero magnetic field $H=0$. If we fit it for small field region,  
medium and large field regions are not fit at all. We believe that at $T=1.9K$ the fermi surface is still not small enough and our theory can not fit such weakly interacting regime by tuning all 4 parameters. 

\begin{figure}[ht!]
\centering
    \subfigure[ ]
   {\includegraphics[width=3.3cm]{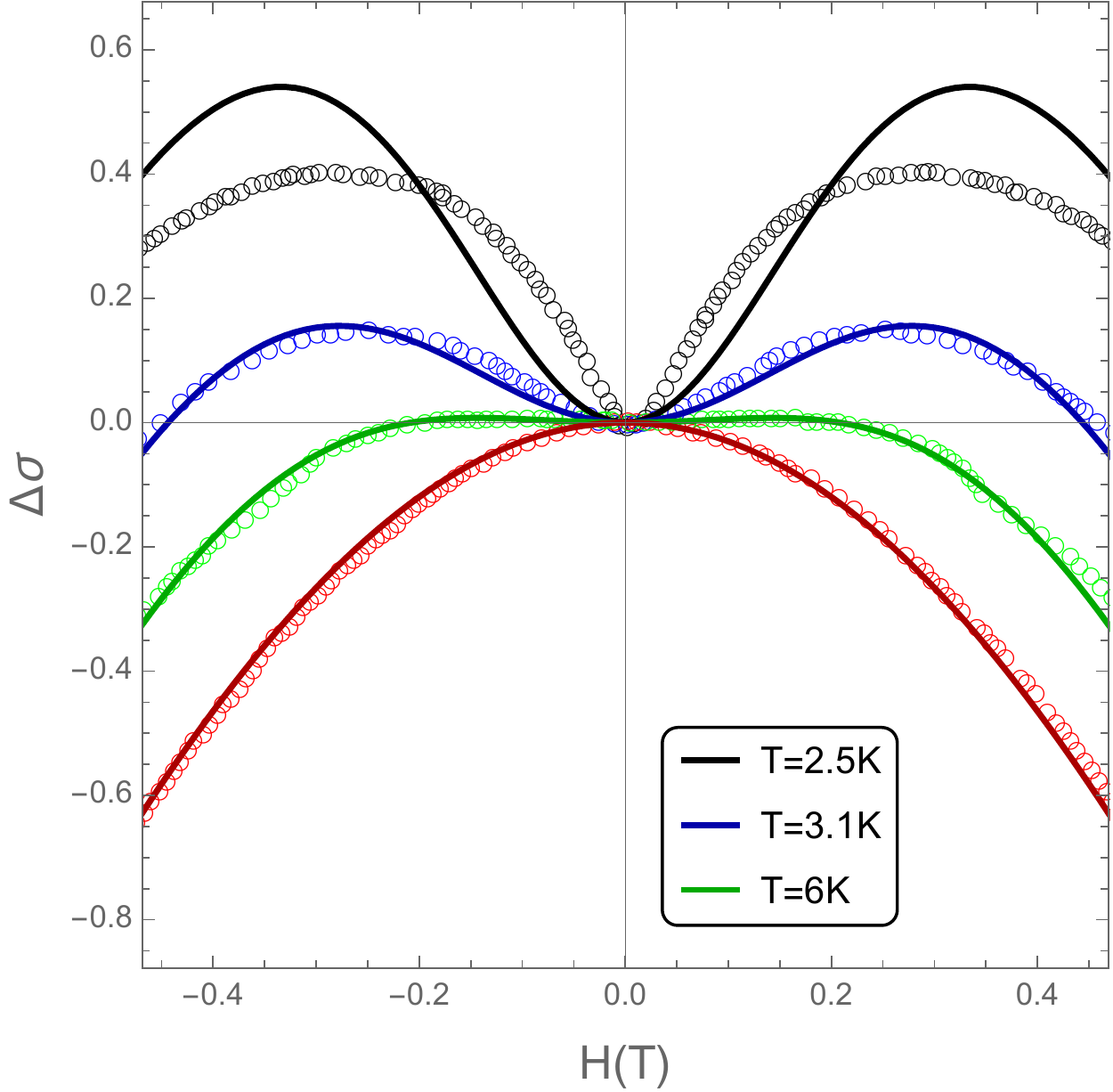}  }
\hskip.5cm   
       \subfigure[ ]
   {\includegraphics[width=4.2cm]{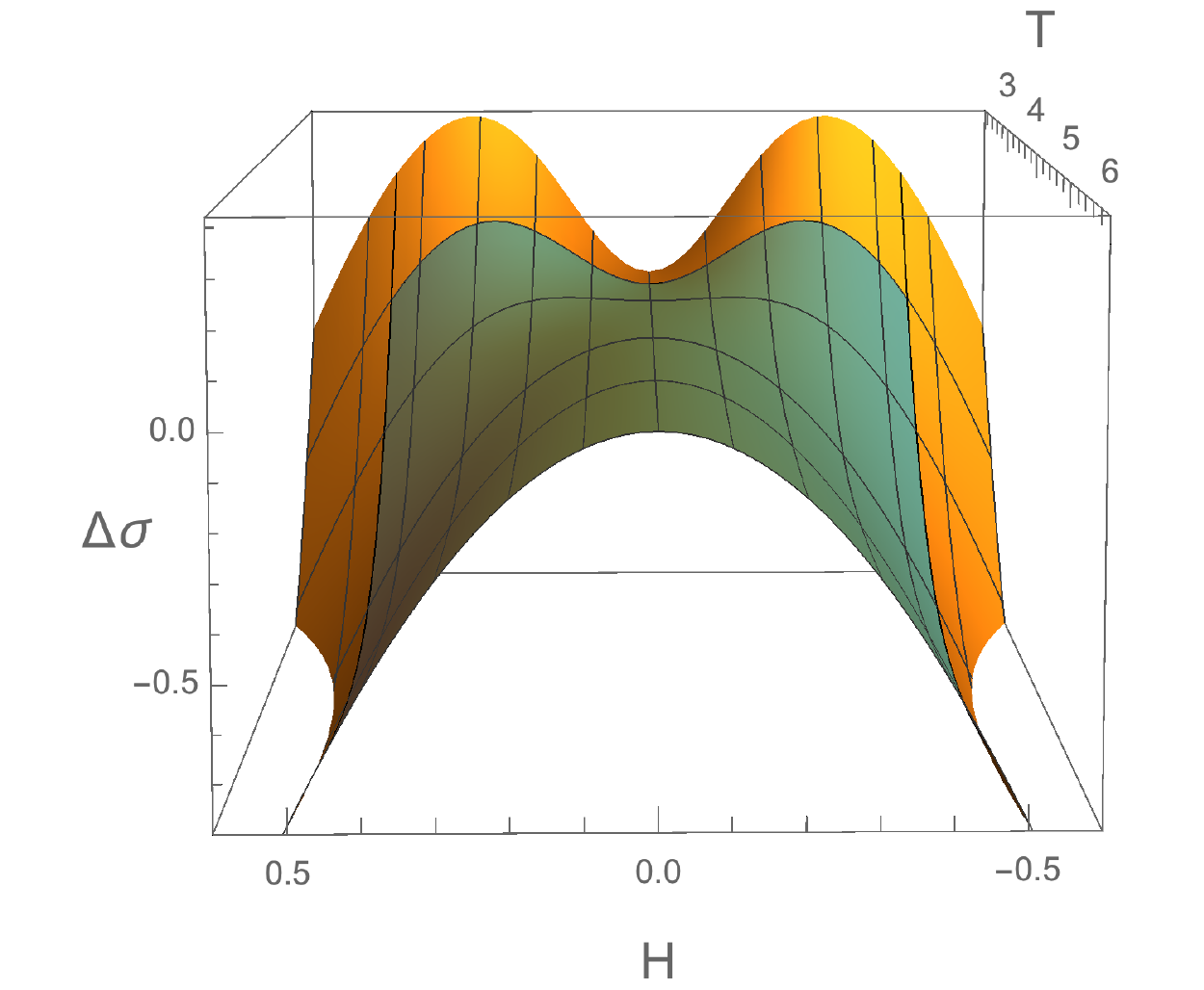}  }
    \caption{ (a) Theory v.s data (circle) for x=0.1.    T=1.9K is in fermi liquid regime where our theory does not work.  (b) $\Delta \sigma$ as function of  $H$  and   $T$.  Our theory works in the green colored island of $(H,T)$ space, where the system is strongly correlated. 
We used  $\beta ^2=\frac{2747}{(\mu m)^2}$, $v_F =7.5 \times 10^4 m/s$, $q_{\chi}=7.12$. 
     } \label{fig:island}
\end{figure}

Another important question is whether our result is universal, namely, independent of details of the matter. 
To answer this question at least partially, we worked out two materials  in the validity islands which is shown 
in figure 3(b).  Figure 3(a) shows  a remarkable similarity in MC curves for different TI material. The transition behavior   seems be universal and well described by our theory. 
\begin{figure}[ht!]
\centering
    \subfigure[ ]
   {\includegraphics[width=3.5cm]{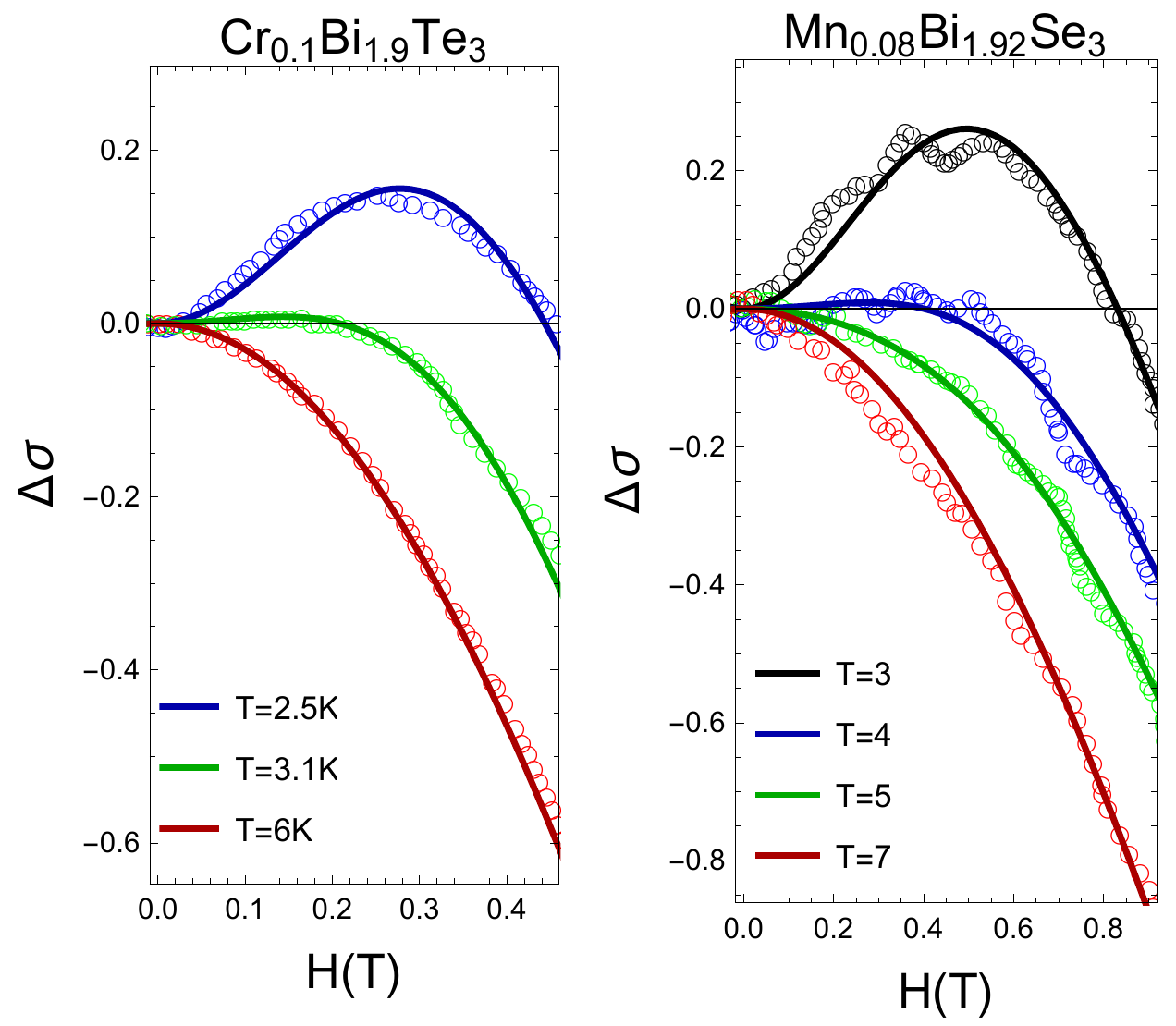}  }
       \subfigure[ ]
   {\includegraphics[width=4.5cm]{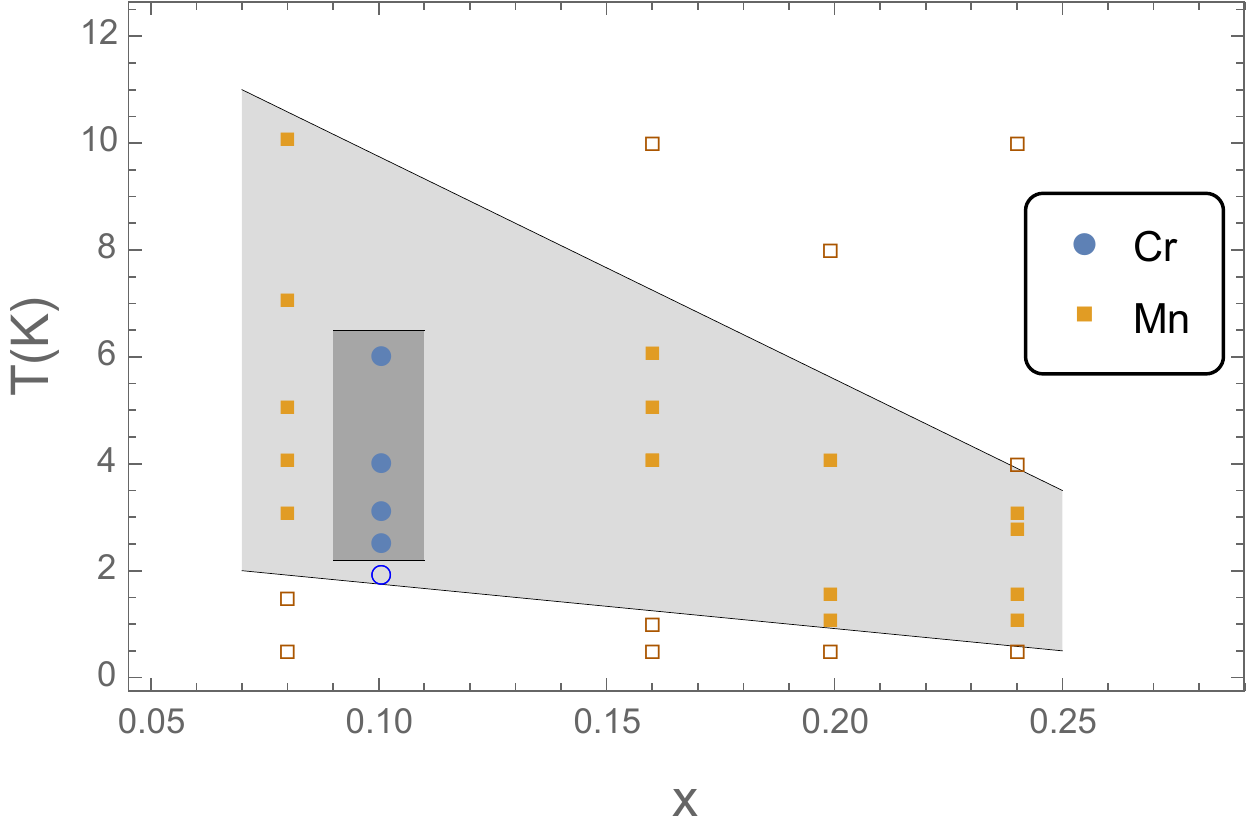}  }
    \caption{ Universality of transition behavior: two different materials are described by the same 
    analytic expression with different parameter values. 
(a)  MC  for Cr doped   $Bi_{2}Te_{3}$ (left)  and      for Mn doped   $Bi_{2}Se_{3}$ (right).  
The data are from ref. [20] and [19] respectively. 
(b) strong correlation islands for the two.  $Bi_{2}Se_{3}$ has bigger 
island due to the bigger bulk gap.  } \label{fig:fitdata}
\end{figure} 
  
In weakly interacting picture, the  non-trivial behavior of magneto-conductivity in crossover regime 
is  understood by the competition between anti-localization induced by 
 spin-orbit coupling  and the localization by surface gap. In holographic picture, 
 the enhancement  in conductivity can be understood as magneto-electric effect 
or  Witten effect. The  interaction term dictates that  external magnetic field generates extra charge carriers $\delta q \sim \theta H$ to increase the conductivity. 
The result of the competition is the sign change in the  curvature of MC curve near $H=0$, where 
\begin{align}
\Delta\sigma \sim  -\frac{2 (1-4\theta^2/9)}{r_0^2 \beta^2} H^2 +{\cal O}(H^4).
\end{align}
and   $\theta = q_{\chi}\gamma \beta^2 /r_0^2$.   
It also explains why    crossover  from WAL to WL appears only in relatively low but not very low temperature region,  because $r_0 \sim T$ for   high temperature  and $\theta$ becomes small so that  $1 -2 \theta/3$ cannot change the sign.  This can be more precisely stated in terms of the phase diagram which is drawn in Fig2(b). Notice that there are only 2 phases. If $\gamma q_{\chi}>1/4$, there is always a phase transition from WAL to WL.  
 
 \vskip.2cm  
{\bf Predictions:} 
Finally we give a list of prediction coming from our theory that can be testable by experiments. 
\begin{itemize}
\item
Near Dirac point of small dopping, we will find transport anomaly, large violation of Wiedemann-Franz Law just like graphene. \item 
For undoped or   weakly doped  TI, 
where one normally see a sharp peak, the characteristic of weak  anti-localization. We predict that if one looks at near Dirac point by adjusting fermi surface by gating for example, 
one  will see the disappearance of the sharp peak as we move down the fermi surface. 
\item 
\vskip-.2cm
We claim that  the transition behavior from WAL $\to$ WL in the medium doping is universal: namely,  
magnetic conductivity of all two dimensional Dirac material with broken TRS, can be described by our formula, 
independent of the detail of the system. 
Here, we  gave only two examples: the Mn doped $Bi_{2}Se_{3}$ in figure 2(a). 
\item 
\vskip-.2cm
For $Cr_{x}Bi_{2-x}Te_{3}$ with $x=0.1$ where  the system in our picture is  strongly interacting for $T\ge 2K$, we expect that ARPES data will show fuzzy density of state(DOS). 
This means that  DOS will be non-zero in the region between dispersion curves,  
where quasi-particle case would show empty DOS leading to the gap. 
\item
\vskip-.2cm
All magneto-transport coefficients other than magneto-conductivity are predictions:  
That is we calculated all the transport coefficients: heat transports thermo-electric power as well as magento-conductance. Once we determine all the coefficients using MC data, all other transport results are predictions. 
It is prediction for several observables as functions of multi-variables $(B, T, \gamma)$, containing a huge set of data.
\end{itemize}
 \vskip -.2cm
{\bf  Future directions}:
In this letter, we  examined the  zero charge sector only.  Nonzero charge parameter $q$ will be discussed in followup paper. 
Other transport coefficients like  thermal conductivities and Seeback coefficients 
with or without magnetic fields are also important aspects that request future investigations.  The graphene has even number of Dirac cones,  weak spin-orbit interaction  and  different  mechanism for WL/WAL. 
Because of such differences, we need to find other  interaction term in holographic model for graphene.
It is also interesting to classify all possible pattern of interaction that provides the fermion surface gap in the presence of strong e-e correlation in our context. 

\vskip 1cm

 \begin{acknowledgements}
 We thank Philip Kim for  comments on the draft  and for stimulating discussions on 
 Dirac fluid. This  work is supported by Mid-career Researcher Program through the National Research Foundation of Korea grant No. NRF-2016R1A2B3007687.  YS is also supported in part by Basic Science Research Program through NRF grant No. NRF-2016R1D1A1B03931443.
\end{acknowledgements}


\bibliography{Refs}

\end{document}